# Initial condition of the string relaxation equation of the string model for glass transition: part-I

J.L. Zhang [a], L.N. Wang [a], J.G. Jiang [b], L.L. Zhang [a], Y.N Huang [a,b]*

*(a) College of Physics and Electronic Information and Institute of Condensed Matter Physics and Design, Ili Normal University, Yining, 835000*
*(b) Department of Physics and National Lab of Solid State Microstructures, Nanjing University, Nanjing 210093*

## Abstract

The string relaxation equation (SRE) of the string model for the glass transition contains the well-known Debye and Rouse-Zimm relaxation equations. However, its initial condition, necessary to the model predictions of glassy dynamics, such as the mechanism of the universal primary α- and Johari-Goldstein β-relaxations in glassformers, has not been solved. In this paper, the special initial condition (SIC) of the SRE of straight strings for dielectric spectrum technique, which is one of the most common methods to measure the glassy dynamics, was tentatively calculated by a direct calculation method, finding that the method has not any practical feasibility. However, a recursive calculation method was developed that allows to obtain the SIC exactly. It should be expected that the obtained SIC would benefit the thorough solution of the general initial condition of the SRE of the string model for stochastically spatially configurating strings, as will be described in separate publications.

## I. Introduction

Though the glass transition is a fundamental property of the condensed matter, none of the mechanisms and models proposed to explain this property has received widely acceptance, remaining a central unresolved issue in condensed matter physics [1-15]. Recently, a molecular level mechanism of coupling dynamical molecule strings was proposed by Huang, Wang, and Riande, called the string model for the sake of convenience [16]. Besides describing the single molecular motions of Debye's mean-field theory approximation [17], the string model represents the collective units of the structure and dynamics in glassformers by coupling molecular strings randomly distributed in space, characteristics which are widely observed in well-designed experiments [18-19], analog simulations [20] and molecular dynamics simulations [21]. Furthermore, the model assumes relative weak interactions between the strings, compared with the intra-string ones.

The Hamiltonian describing orientational motions in Debye's theory [17], is the well-known sin-Gordon potential, $H_0 = \frac{V_0}{2} \sum_i \{1 - \cos[2(\varphi_i - \varphi_i^0)]\}$, where $V_0$ is the barrier height between the double-wells, $\phi_i$ ($0 \leq \phi_i < \pi$) is the orientational angle of the i[th] molecule in the system, and $\phi_i^0$ is the referenced angle in the range $[0, \pi]$. Translational movements are described by the $\phi^{(4)}$ potential, $H_0 = V_0 \sum_i \left[ -2\left(\frac{x_i}{d}\right)^2 + \left(\frac{x_i}{d}\right)^4 \right]$, where $V_0$, $x_i$ and $d$ are, respectively, the barrier height between the double-wells, the displacement of i[th] molecule in the system and the well width.

---

* Author to whom correspondence should be addressed. Email: ynhuang@nju.edu.cn

Furthermore, molecular jumping processes between the double-wells ($e^{-V_0/k_B T} \ll 1$, where $k_B$ is the Boltzmann's constant, and $T$ the absolute temperature) are equivalent to those between two states, $\sigma_i = \pm 1$, with jumping rate $v_0 e^{-V_0/k_B T}$ ($v_0$ is the vibration frequency of molecules in the wells). Taking into account these principles, the string model represents intra-string interactions by a finite one dimensional Ising model $H_1 = -V \sum_m \sum_{i=1}^{n-1} \sigma_i^{mn} \sigma_{i+1}^{mn}$ [17], whereas the inter-string interactions are represented by a random Ising interacting model $H_2 = \frac{V'}{2} \sum_m \sum_{i=1}^{n} \sum_{m' \neq m, i'}^{nn(i)} \sigma_i^{mn} \sigma_{i'}^{m'n'} \Theta_{ii'}^{mm'}$, where $V$ and $V'$ are the corresponding interaction constants, independent of temperature. The symbol $\sigma_i^{mn} = \pm 1$ denotes the two states of the i$^{th}$ molecule of a string (numbered $m$ in the system) containing $n$ molecules (called $n$-string hereafter), $nn(k)$ represents the nearest number of molecules surrounding the i$^{th}$ molecule, and $\Theta_{ii'}^{mm'}$ is a random number in the range $[-1, 1]$. In the case of orientational motions, $\Theta_{ii'}^{mm'}$ is the direction cosine of the molecule $i$ in the $n$-string with the molecule $i'$ in the $n'$-string. For the Debye's theory only provides a jumping time for the molecules between the double-wells $\tau_0 = v_0^{-1} e^{V_0/k_B T}$, the effective Hamiltonian of the string model describing the coupled orientational and translational jumping motions is given by,

$$H_e = -V \sum_m \sum_{i=1}^{n-1} \sigma_i^{mn} \sigma_{i+1}^{mn} + \frac{V'}{2} \sum_m \sum_{i=1}^{n} \sum_{m' \neq m, i'}^{nn(i)} \sigma_i^{mn} \sigma_{i'}^{m'n'} \Theta_{ii'}^{mm'} \qquad (1)$$

This equation is the Hamiltonian of a partly random Ising model [17].

The main conclusions of the model for a glassy system under a first-order approximation (the calculation error is estimated to be about 1% near the glass transition temperature $T_g$) are [16]: 1) the relaxation behavior of a molecular string is equivalent to that of an effective molecule (EM), characterized by a definite relaxation time and a dipole moment; 2) an ordinary liquid at high temperature is renormalized to an EM gas, this approach presenting a possible brand-new kind picture about the ordinary liquid; 3) the strong correlated supercooled liquid at low temperature is renormalized to a weak correlated EM liquid; and 4) there exists a crossover transition between the EM gas and the EM liquid. The polarizability of the system predicted by the model gives a unified and quantitative description of the α-relaxation dynamics in glass formers including: i) the relaxation time crossover from high temperature Arrhenius behavior to low temperature Vogel-Fulcher-Tammann law [22]; ii) the crossover of the relaxation function from a high temperature Debye exponential to a low temperature Kohlrausch-Williams-Watts [23] or Cole-Davidson [24] functions; and iii) the departure of the relaxation intensity form the Curie law [13-15]. In addition, comparing with other models, least parameters of clear physical origin are involved in the string model [16].

The mathematical core of the string model arises from the intra-string interaction $H_1$ described by Eq.(1), which leads to the following relaxation equations for an individual n-string [16],

$$\frac{d}{dt}\begin{bmatrix} \delta_1 \\ \delta_2 \\ \vdots \\ \vdots \\ \vdots \\ \delta_{n-1} \\ \delta_n \end{bmatrix} = -v_0 e^{-V_0/kT} \begin{bmatrix} 1 & 2u-1 & 0 & \cdots & 0 & 0 & 0 \\ w-1/2 & 1 & w-1/2 & \cdots & \vdots & 0 & 0 \\ \vdots & w-1/2 & 1 & \cdots & \vdots & \vdots & 0 \\ \vdots & \vdots & w-1/2 & \cdots & w-1/2 & \vdots & \vdots \\ 0 & \vdots & \vdots & \cdots & 1 & w-1/2 & \vdots \\ 0 & 0 & \vdots & \cdots & w-1/2 & 1 & w-1/2 \\ 0 & 0 & 0 & \cdots & 0 & 2u-1 & 1 \end{bmatrix} \begin{bmatrix} \delta_1 \\ \delta_2 \\ \vdots \\ \vdots \\ \vdots \\ \delta_{n-1} \\ \delta_n \end{bmatrix} \quad (2)$$

where $\delta_i^{(n)}(t) \equiv p_i^{(n)}(t) - p_i^{(n)}(\infty), i=1,\cdots,n$, $p_i^{(n)}(t)$ is the probability that the i$^{th}$ molecule of the n-string is in the orientational state $\sigma_i^{mn}=1$, and $p_i^{(n)}(\infty)$ is the relevant probability at equilibrium. Therefore, $\delta_i^{(n)}(t)$ is the departure of the property from equilibrium. $u \equiv (1+e^{2v})^{-1}$, $w \equiv (1+e^{v})^{-1}$, $v \equiv V/k_B T$, $k_B$ is the Boltzmann constant, and $T$ is the absolute temperature of the system.

For the sake of convenience, Eq.(2) will be called the string relaxation equation (SRE) whereas the square matrix on the right side of Eq.(2) will be named $M_{SR}$. At high enough temperature, i.e. $v \to 0$, $M_{SR}$ becomes a unit matrix and the SRE simplifies to the Debye relaxation equation [17]. On the other hand, at low enough temperature, i.e. $v \to \infty$, and $M_{SR}$ is the Rouse-Zimm matrix, the SRE becoming the Rouse-Zimm relaxation equation [25]. Since the Debye theory and the Rouse model are the two most successful relaxation theories for monomers and macromolecules, respectively, the string model can be considered, at least mathematically a universal model that describes the relaxation dynamics of the amorphous condensed matter.

Physically, the relaxation of the n coupled molecules in an n-string is described by n individual relaxation modes, according to the SRE. The first-order approximation in Ref.16 only gives the main relaxation mode corresponding to the largest relaxation intensity and the longest relaxation time, the other modes being omitted. In addition to the α-relaxation, glass formers exhibit a universal well-known Johari-Goldstein β-relaxation at high frequencies [13-15]. The microscopic mechanism at molecular level associated with the Johari-Goldstein β-relaxation is not clear till now though there are some phenomenological theories, such as the coupling model [7] and the mode-coupling theory [8], which give a glimpse of the characteristics of this process. The relaxation mode scenario of the string model possibly provides a unified mechanism of the α- and β-relaxations. It can be assumed that with the exception of the main relaxation mode associated to the α-relaxation, all the other modes correspond to the Johari-Goldstein β-process. To carry out the exact analysis and quantitative comparison of these assumptions with experiments, it is necessary to estimate the strengths of all the modes. To accomplish this goal, it is necessary to solve the initial conditions of the SRE, that is,

$$\delta_i^{(n)}(0) \equiv p_i^{(n)}(0) - p_i^{(n)}(\infty), \; i=1,\cdots,n \quad (3)$$

Owing to the stochastic spatial configurations of the molecular strings, the general solution of the initial condition involves serious difficulties. Because of this, in what follows, we only study the case of straight molecular strings, a special case, called SIC of the SRE.

## 2. General calculation method

Similar to the method described in Ref.16, the relevant question of the motion of n molecules coupled in an n-string is first transformed to that of the individual $2^n$ orientational states. The motion of each one of these $2^n$ states can be exactly calculated using the Boltzmann principle for the individual states, and the results thus obtained are transformed back to those of the coupled molecules.

Although the SRE is independent of experimental methods in the linear response regime, its initial condition shown as Eq.(3) depends. In Ref.16, Huang, Wang and Riande have calculated the permittivity of glass formers because this property provides the glassy dynamics in a wide frequency range [13-15]. In this technique, the relevant SIC of an n-string corresponding to $\delta_i^{(n)}(0), i=1,\cdots,n$ is induced by an external electric field. Without losing generality, and in the linear response regime [17], let us consider an n-string perturbed by a small enough step electric field along the string,

$$F(t) = \begin{cases} F_0, & t < 0 \\ 0, & t \geq 0 \end{cases} \quad (4)$$

Mathematically, the linear response condition is that the reduced energy of the molecule in the electric field $f \equiv \mu_0 F_0 / k_B T \ll 1$. In this expression, $\mu_0$ is the permanent electric dipole moment of the molecule along the string direction.

## 3. Direct calculation method of SIC and questions

Since a molecule in the string model has two possible orientations, an n-string has $2^n$ orientational states. Let $E_j^{(n)}(F)$ and $q_j^{(n)}(t)$ be, respectively, the energy and the probability of the j[th] orientational state of the n-string in the electric field. According to the Boltzmann principle,

$$\begin{cases} q_j^{(n)}(0) = e^{-E_j^{(n)}(F_0)/kT} \Big/ \sum_{k=1}^{2^n} e^{-E_k^{(n)}(F_0)/kT} \\ q_j^{(n)}(\infty) = e^{-E_j^{(n)}(0)/kT} \Big/ \sum_{k=1}^{2^n} e^{-E_k^{(n)}(0)/kT} \end{cases} \quad (5)$$

As an example, the $2^4$ orientational states of a 4-string are shown in table-1, where $Q_n = 2(2chv)^{n-1}$ is the orientational partition function of the n-string [16], and $chv \equiv (e^v + e^{-v})/2$. Since the two orientations $\sigma_i^{mn} = 1$ and $\sigma_i^{mn} = -1$ of a molecule can be formally expressed by "0" and "1", respectively, the total orientational states of an n-string can be denoted by n binary digits, as shown in the 2[nd] columns of table-1. For example, "0000" indicates that $\sigma_i^{m4} = 1, i = 1,2,3,4$. The 3[rd] and 4[th] columns denote $E_j^{(4)}(0)$ and $E_j^{(4)}(F_0)$ for the j[th] orientational state when the external field is 0 and $F_0$, respectively, whereas the relevant quantities $q_j^{(4)}(\infty)$ and $q_j^{(4)}(0)$ for $t \to \infty$ and $t = 0$, respectively, are shown in the 5th and 6th columns. $E_j^{(n)}(F)$ has two parts, one originating from the intra-string interaction $H_1$ of Eq.(1) and the other from the permanent electric dipole moment of the molecule in the electric field, e.g. for the "0000" orientational state, the intra-string energy and the energy of string molecules in the electric field are $-3V$ and $-4\mu_0 F$, respectively. For the linear response case $f \ll 1$ and $e^f \approx 1+f$, as shown in the 6th column of the table-1.

Table-1 Expressions for the total orientational states of a straight 4-string and the relevant quantities in the external electric field of Eq.(4)

| serial number | orientational states | $E_j^{(4)}(0)$ | $E_j^{(4)}(F_0)$ | $q_j^{(4)}(\infty)Q_4$ | $q_j^{(4)}(0)Q_4$ |
|---|---|---|---|---|---|
| $j=1$ | 0000 | $-3V$ | $-3V - 4\mu_0 F_0$ | $e^{3v}$ | $(1+4f)e^{3v}$ |
| $j=2$ | 0001 | $-V$ | $-V - 2\mu_0 F_0$ | $e^{v}$ | $(1+2f)e^{v}$ |
| $j=3$ | 0010 | $V$ | $V - 2\mu_0 F_0$ | $e^{-v}$ | $(1+2f)e^{-v}$ |
| $j=4$ | 0011 | $-V$ | $-V$ | $e^{v}$ | $e^{v}$ |
| $j=5$ | 0100 | $V$ | $V - 2\mu_0 F_0$ | $e^{-v}$ | $(1+2f)e^{-v}$ |
| $j=6$ | 0101 | $3V$ | $3V$ | $e^{-3v}$ | $e^{-3v}$ |
| $j=7$ | 0110 | $V$ | $V$ | $e^{-v}$ | $e^{-v}$ |
| $j=8$ | 0111 | $-V$ | $-V + 2\mu_0 F_0$ | $e^{v}$ | $(1-2f)e^{v}$ |
| $j=9$ | 1000 | $-V$ | $-V - 2\mu_0 F_0$ | $e^{v}$ | $(1+2f)e^{v}$ |
| $j=10$ | 1001 | $V$ | $V$ | $e^{-v}$ | $e^{-v}$ |
| $j=11$ | 1010 | $3V$ | $3V$ | $e^{-3v}$ | $e^{-3v}$ |
| $j=12$ | 1011 | $V$ | $V + 2\mu_0 F_0$ | $e^{-v}$ | $(1-2f)e^{-v}$ |
| $j=13$ | 1100 | $-V$ | $-V$ | $e^{v}$ | $e^{v}$ |
| $j=14$ | 1101 | $V$ | $V + 2\mu_0 F_0$ | $e^{-v}$ | $(1-2f)e^{-v}$ |
| $j=15$ | 1110 | $-V$ | $-V + 2\mu_0 F_0$ | $e^{v}$ | $(1-2f)e^{v}$ |
| $j=16$ | 1111 | $-3V$ | $-3V + 4\mu_0 F_0$ | $e^{3v}$ | $(1-4f)e^{3v}$ |

From the definition given for $p_i^{(n)}(t)$ above and taking into account table-1, the following relationships between $p_i^{(4)}(t)$ and $q_j^{(4)}(t)$ are obtained,

$$\begin{cases} p_1^{(4)}(t) = q_1^{(4)}(t) + q_2^{(4)}(t) + q_3^{(4)}(t) + q_4^{(4)}(t) + q_5^{(4)}(t) + q_6^{(4)}(t) + q_7^{(4)}(t) + q_8^{(4)}(t) \\ p_2^{(4)}(t) = q_1^{(4)}(t) + q_2^{(4)}(t) + q_3^{(4)}(t) + q_4^{(4)}(t) + q_9^{(4)}(t) + q_{10}^{(4)}(t) + q_{11}^{(4)}(t) + q_{12}^{(4)}(t) \\ p_3^{(4)}(t) = q_1^{(4)}(t) + q_2^{(4)}(t) + q_5^{(4)}(t) + q_6^{(4)}(t) + q_9^{(4)}(t) + q_{10}^{(4)}(t) + q_{13}^{(4)}(t) + q_{14}^{(4)}(t) \\ p_4^{(4)}(t) = q_1^{(4)}(t) + q_3^{(4)}(t) + q_5^{(4)}(t) + q_7^{(4)}(t) + q_9^{(4)}(t) + q_{11}^{(4)}(t) + q_{13}^{(4)}(t) + q_{15}^{(4)}(t) \end{cases} \qquad (6)$$

The specific method for the calculation is to find first all the orientationl state with the first (from left to right) binary digit being "0", as shown in table-1, and then to add them. Thus $\delta_i^{(4)}(0), i=1,\cdots,4$ can exactly be calculated by means of Eqs.(3) and (6) as well as the results of $q_j^{(4)}(0)$ and $q_j^{(4)}(\infty)$ presented in table-1. This procedure is called the direct calculation method.

For an arbitrary string length, the SIC can be calculated by the same method as that used for the 4-string, and the pertinent results for n=1 to 7 are shown in table-2. The results of the table indicate that $\delta_i^{(n)}(0) = \delta_{n-i+1}^{(n)}(0), i=1,\cdots,n$, an equality which could be forecasted from the symmetry of the straight string.

Table-2 Strict solution of the SICSRE for n=1 to 7 in the external field of Eq.(4) using the direct calculation method

| n | $\delta_1^{(n)}(0)Q_n/f$ | $\delta_2^{(n)}(0)Q_n/f$ | $\delta_3^{(n)}(0)Q_n/f$ | $\delta_4^{(n)}(0)Q_n/f$ |
|---|---|---|---|---|
| 1 | 1 | | | |
| 2 | $2e^v$ | | | |
| 3 | $3e^{2v} + e^{-2v}$ | $3e^{2v} + 2 - e^{-2v}$ | | |
| 4 | $4e^{3v} + 4e^{-v}$ | $4e^{3v} + 4e^v$ | | |
| 5 | $5e^{4v} + 10 + e^{-4v}$ | $5e^{4v} + 6e^{2v} + 4 + 2e^{-2v} - e^{-4v}$ | $5e^{4v} + 8e^{2v} + 2 + e^{-4v}$ | |
| 6 | $6e^{5v} + 20e^v + 6e^{-3v}$ | $6e^{5v} + 8e^{3v} + 12e^v + 8e^{-v} - 2e^{-3v}$ | $6e^{5v} + 12e^{3v} + 8e^v + 4e^{-v} + 2e^{-3v}$ | |
| 7 | $7e^{6v} + 36e^{2v} + 20e^{-4v} + e^{-6v}$ | $7e^{6v} + 10e^{4v} + 25e^{2v} + 20 + 2e^{-2v} + 2e^{-4v} - e^{-6v}$ | $7e^{6v} + 10e^{4v} + 25e^{2v} + 20 + 2e^{-2v} + 2e^{-4v} - e^{-6v}$ | $7e^{6v} + 18e^{4v} + 18e^{2v} + 18 + 8e^{-2v} + 2e^{-4v} - e^{-6v}$ |

An inspection of the results of table-2 shows that in increasing n, the expressions for $\delta_i^{(n)}(0)$ become more and more complex. Specifically, the number of terms for $\delta_i^{(n)}(0)$ increases rapidly and the exponential factors of each term become higher and higher. Unfortunately, except for the first term, the regularity of $\delta_i^{(n)}(0)$ for arbitrary values of n is not found. We would like to point out that for n ~ 50, the number of total orientational states of an n-string is roughly $2^{50}$ around the glass transition temperature [16]. Then the use of the direct calculation method is not practically feasible.

## 4. Recursive calculation method for SIC

This section describes a recursive method to calculate $\delta_i^{(n)}(0)$. The key point of the method is to relate the energy of $j^{th}$ orientational state of the n-string $E_j^{(n)}(F)$, to the energies of shorter strings $E_k^{(n-1)}(F)$, $E_l^{(n-2)}(F)$, $\cdots$, $E_1^{(1)}(F)$ in the external field. Then the recursive relation between $q_j^{(n)}(t)$ and $q_k^{(n-1)}(t)$, $q_l^{(n-2)}(t)$, $\cdots$ $q_1^{(1)}(t)$, and consequently the recursive relations between $p_i^{(n)}(0)$ of the n-string and $p_k^{(n-1)}(t)$, $p_l^{(n-2)}(t)$, $\cdots$, $p_1^{(1)}(t)$ of shorter strings are found. Finally, the values of $\delta_i^{(n)}(0)$, $i=1,\cdots,n$ are calculated using Eq.(3). In what follows, the 4-string is utilized as an example to calculate $p_i^{(n)}(\infty)$. The specific calculation processes are,

$$
\begin{aligned}
p_1^{(4)}(\infty) &= q_1^{(4)}(\infty) + q_2^{(4)}(\infty) + q_3^{(4)}(\infty) + q_4^{(4)}(\infty) + q_5^{(4)}(\infty) + q_6^{(4)}(\infty) + q_7^{(4)}(\infty) + q_8^{(4)}(\infty) \\
&= \left[ e^{-e_1^{(4)}(0)} + e^{-e_2^{(4)}(0)} + e^{-e_3^{(4)}(0)} + e^{-e_4^{(4)}(0)} + e^{-e_5^{(4)}(0)} + e^{-e_6^{(4)}(0)} + e^{-e_7^{(4)}(0)} + e^{-e_8^{(4)}(0)} \right] / Q_4 \\
&= \left[ e^{-e_1^{(3)}(0)+v} + e^{-e_2^{(3)}(0)+v} + e^{-e_3^{(3)}(0)+v} + e^{-e_4^{(3)}(0)+v} + e^{-e_5^{(3)}(0)-v} + e^{-e_6^{(3)}(0)-v} + e^{-e_7^{(3)}(0)-v} + e^{-e_8^{(3)}(0)-v} \right] / Q_4 , \\
&= \left[ \left( q_1^{(3)}(\infty) + q_2^{(3)}(\infty) + q_3^{(3)}(\infty) + q_4^{(3)}(\infty) \right) e^{v} + \left( q_5^{(2)}(\infty) + q_6^{(2)}(\infty) + q_7^{(3)}(\infty) + q_8^{(3)}(\infty) \right) e^{-v} \right] Q_3 / Q_4 \\
&= \left[ p_1^{(3)}(\infty) e^{v} + \left(1 - p_1^{(3)}(\infty)\right) e^{-v} \right] Q_3 / Q_4
\end{aligned}
$$

$$
\begin{aligned}
p_2^{(4)}(\infty) &= q_1^{(4)}(\infty) + q_2^{(4)}(\infty) + q_3^{(4)}(\infty) + q_4^{(4)}(\infty) + q_9^{(4)}(\infty) + q_{10}^{(4)}(\infty) + q_{11}^{(4)}(\infty) + q_{12}^{(4)}(\infty) \\
&= \left[ e^{-e_1^{(4)}(0)} + e^{-e_2^{(4)}(0)} + e^{-e_3^{(4)}(0)} + e^{-e_4^{(4)}(0)} + e^{-e_9^{(4)}(0)} + e^{-e_{10}^{(4)}(0)} + e^{-e_{11}^{(4)}(0)} + e^{-e_{12}^{(4)}(0)} \right] / Q_4 \\
&= \left[ e^{-e_1^{(3)}(0)+v} + e^{-e_2^{(3)}(0)+v} + e^{-e_3^{(3)}(0)+v} + e^{-e_4^{(3)}(0)+v} + e^{-e_1^{(3)}(0)-v} + e^{-e_2^{(3)}(0)-v} + e^{-e_3^{(3)}(0)-v} + e^{-e_4^{(3)}(0)-v} \right] / Q_4 , \\
&= \left( q_1^{(3)}(\infty) + q_2^{(3)}(\infty) + q_3^{(3)}(\infty) + q_4^{(3)}(\infty) \right)\left( e^{v} + e^{-v} \right) Q_3 / Q_4 \\
&= p_1^{(3)}(\infty)\left( e^{v} + e^{-v} \right) Q_3 / Q_4
\end{aligned}
$$

$$
\begin{aligned}
p_3^{(4)}(\infty) &= q_1^{(4)}(\infty) + q_2^{(4)}(\infty) + q_5^{(4)}(\infty) + q_6^{(4)}(\infty) + q_9^{(4)}(\infty) + q_{10}^{(4)}(\infty) + q_{13}^{(4)}(\infty) + q_{14}^{(4)}(\infty) \\
&= \left[ e^{-e_1^{(4)}(0)} + e^{-e_2^{(4)}(0)} + e^{-e_5^{(4)}(0)} + e^{-e_6^{(4)}(0)} + e^{-e_9^{(4)}(0)} + e^{-e_{10}^{(4)}(0)} + e^{-e_{13}^{(4)}(0)} + e^{-e_{14}^{(4)}(0)} \right] / Q_4 \\
&= \left[ e^{-e_1^{(3)}(0)+v} + e^{-e_2^{(3)}(0)+v} + e^{-e_5^{(3)}(0)-v} + e^{-e_6^{(3)}(0)-v} + e^{-e_1^{(3)}(0)-v} + e^{-e_2^{(3)}(0)-v} + e^{-e_5^{(3)}(0)+v} + e^{-e_6^{(3)}(0)+v} \right] / Q_4 , \\
&= \left( q_1^{(3)}(\infty) + q_2^{(3)}(\infty) + q_5^{(3)}(\infty) + q_6^{(3)}(\infty) \right)\left( e^{v} + e^{-v} \right) Q_3 / Q_4 \\
&= p_2^{(3)}(\infty)\left( e^{v} + e^{-v} \right) Q_3 / Q_4
\end{aligned}
$$

and

$$p_4^{(4)}(\infty) = q_1^{(4)}(\infty) + q_3^{(4)}(\infty) + q_5^{(4)}(\infty) + q_7^{(4)}(\infty) + q_9^{(4)}(\infty) + q_{11}^{(4)}(\infty) + q_{13}^{(4)}(\infty) + q_{15}^{(4)}(\infty)$$

$$= \left[ e^{-e_1^{(4)}(0)} + e^{-e_3^{(4)}(0)} + e^{-e_5^{(4)}(0)} + e^{-e_7^{(4)}(0)} + e^{-e_9^{(4)}(0)} + e^{-e_{11}^{(4)}(0)} + e^{-e_{13}^{(4)}(0)} + e^{-e_{15}^{(4)}(0)} \right] / Q_4$$

$$= \left[ e^{-e_1^{(3)}(0)+v} + e^{-e_3^{(3)}(0)+v} + e^{-e_5^{(3)}(0)-v} + e^{-e_7^{(3)}(0)-v} + e^{-e_1^{(3)}(0)-v} + e^{-e_3^{(3)}(0)-v} + e^{-e_5^{(3)}(0)+v} + e^{-e_7^{(3)}(0)+v} \right] / Q_4 ,$$

$$= \left( q_1^{(3)}(\infty) + q_3^{(3)}(\infty) + q_5^{(3)}(\infty) + q_7^{(3)}(\infty) \right) \left( e^v + e^{-v} \right) Q_3 / Q_4$$

$$= p_3^{(3)}(\infty)\left( e^v + e^{-v} \right) Q_3 / Q_4$$

where $e_j^{(n)}(F) \equiv E_j^{(n)}(F)/k_B T$, whereas the expressions for $p_i^{(n)}(\infty)$ of the n-string ( n=1 to 4) are shown in table-3.

Table-3 $p_i^{(n)}(\infty)$ of the n-string for n=1 to 4 in the external field of Eq.(4)

| n | $p_1^{(n)}(\infty)Q_n/Q_{n-1}$ | $p_2^{(n)}(\infty)Q_n/Q_{n-1}$ | $p_3^{(n)}(\infty)Q_n/Q_{n-1}$ | $p_4^{(n)}(\infty)Q_n/Q_{n-1}$ |
|---|---|---|---|---|
| 1 | 1/2 | | | |
| 2 | $p_1^{(1)}(\infty)e^v + \left(1 - p_1^{(1)}(\infty)\right)e^{-v}$ | $p_1^{(1)}(\infty)\left(e^v + e^{-v}\right)$ | | |
| 3 | $p_1^{(2)}(\infty)e^v + \left(1 - p_1^{(2)}(\infty)\right)e^{-v}$ | $p_1^{(2)}(\infty)\left(e^v + e^{-v}\right)$ | $p_2^{(2)}(\infty)\left(e^v + e^{-v}\right)$ | |
| 4 | $p_1^{(3)}(\infty)e^v + \left(1 - p_1^{(3)}(\infty)\right)e^{-v}$ | $p_1^{(3)}(\infty)\left(e^v + e^{-v}\right)$ | $p_2^{(3)}(\infty)\left(e^v + e^{-v}\right)$ | $p_3^{(3)}(\infty)\left(e^v + e^{-v}\right)$ |

From table-3, the recursive relation of $p_i^{(n)}(\infty)$ for arbitrary n is,

$$\begin{cases} p_1^{(n)}(\infty) = \left[ p_1^{(n-1)}(\infty)e^v + \left(1 - p_1^{(n-1)}(\infty)\right)e^{-v} \right] Q_{n-1}/Q_n , \ n \geq 2 \\ p_i^{(n)}(\infty) = p_{i-1}^{(n-1)}(\infty)\left(e^v + e^{-v}\right) Q_{n-1}/Q_n , \ n \geq 2, i = 2,\cdots,n \end{cases} \quad (7)$$

Eq.(7) readily gives $p_i^{(n)}(\infty) = 1/2$, and in fact the above results can easily be obtained from the symmetry of the straight strings. The main purpose of using the methodology described above to reach Eq.(7) is to present useful mathematical methods to deduce the more complicated $p_i^{(n)}(0)$ parameter.

By taking a 4-string as an example, the calculation processes involved in the estimation of $p_i^{(n)}(0)$ are,

$$p_1^{(4)}(0) = q_1^{(4)}(0) + q_2^{(4)}(0) + q_3^{(4)}(0) + q_4^{(4)}(0) + q_5^{(4)}(0) + q_6^{(4)}(0) + q_7^{(4)}(0) + q_8^{(4)}(0)$$

$$= \left[ e^{-e_1^{(4)}(F_0)} + e^{-e_2^{(4)}(F_0)} + e^{-e_3^{(4)}(F_0)} + e^{-e_4^{(4)}(F_0)} + e^{-e_5^{(4)}(F_0)} + e^{-e_6^{(4)}(F_0)} + e^{-e_7^{(4)}(F_0)} + e^{-e_8^{(4)}(F_0)} \right] / Q_4$$

$$= \left[ \begin{array}{l} e^{-e_1^{(3)}(F_0)+v+f} + e^{-e_2^{(3)}(F_0)+v+f} + e^{-e_3^{(3)}(F_0)+v+f} + e^{-e_4^{(3)}(F_0)+v+f} \\ + e^{-e_5^{(3)}(F_0)-v+f} + e^{-e_6^{(3)}(F_0)-v+f} + e^{-e_7^{(3)}(F_0)-v+f} + e^{-e_8^{(3)}(F_0)-v+f} \end{array} \right] / Q_4 \quad (8)$$

$$= \left[ \left( q_1^{(3)}(0) + q_2^{(3)}(0) + q_3^{(3)}(0) + q_4^{(3)}(0) \right) e^{v} + \left( q_5^{(2)}(0) + q_6^{(2)}(0) + q_7^{(3)}(0) + q_8^{(3)}(0) \right) e^{-v} \right] e^{f} Q_3 / Q_4$$

$$\approx \left[ p_1^{(3)}(0) e^{v} + \left( 1 - p_1^{(3)}(0) \right) e^{-v} \right] (1+f) Q_3 / Q_4$$

$$p_2^{(4)}(0) = q_1^{(4)}(0) + q_2^{(4)}(0) + q_3^{(4)}(0) + q_4^{(4)}(0) + q_9^{(4)}(0) + q_{10}^{(4)}(0) + q_{11}^{(4)}(0) + q_{12}^{(4)}(0)$$

$$= \left[ e^{-e_1^{(4)}(F_0)} + e^{-e_2^{(4)}(F_0)} + e^{-e_3^{(4)}(F_0)} + e^{-e_4^{(4)}(F_0)} + e^{-e_9^{(4)}(F_0)} + e^{-e_{10}^{(4)}(F_0)} + e^{-e_{11}^{(4)}(F_0)} + e^{-e_{12}^{(4)}(F_0)} \right] / Q_4$$

$$= \left[ \begin{array}{l} e^{-e_1^{(3)}(F_0)+v+f} + e^{-e_2^{(3)}(F_0)+v+f} + e^{-e_3^{(3)}(F_0)+v+f} + e^{-e_4^{(3)}(F_0)+v+f} \\ + e^{-e_1^{(3)}(F_0)-v-f} + e^{-e_2^{(3)}(F_0)-v-f} + e^{-e_3^{(3)}(F_0)-v-f} + e^{-e_4^{(3)}(F_0)-v-f} \end{array} \right] / Q_4 \quad (9)$$

$$= \left( q_1^{(3)}(0) + q_2^{(3)}(0) + q_3^{(3)}(0) + q_4^{(3)}(0) \right) \left( e^{v+f} + e^{v-f} \right) Q_3 / Q_4$$

$$\approx p_1^{(3)}(0) \left( 2chv + 2shv \cdot f \right) Q_3 / Q_4$$

where $shv \equiv (e^{v} - e^{-v})/2$.

Both the renumbering of the orientational states shown in table-1 and the symmetry of the straight strings show that $p_i^{(n)}(0) = p_{n-i+1}^{(n)}(0)$. Table-4 presents the following quantities calculated from n=1 to 13 using the method indicated above. $\bar{p}_1^{(n)}(0) \equiv p_1^{(n)}(0) / \left[ p_1^{(n-1)}(0)(1+f) thv \right]$, $\bar{p}_2^{(n)}(0) \equiv \left( p_2^{(n)}(0) / p_1^{(n-1)}(0) - 1 \right) / thv \cdot f$, $\bar{p}_3^{(n)}(0) \equiv \left( p_3^{(n)}(0) / p_1^{(n-2)}(0) - 1 \right) / thv \cdot f$, $\bar{p}_4^{(n)}(0) \equiv \left( p_4^{(n)}(0) / p_1^{(n-3)}(0) - 1 \right) / thv \cdot f$, $\bar{p}_5^{(n)}(0) \equiv \left( p_5^{(n)}(0) / p_1^{(n-4)}(0) - 1 \right) / thv \cdot f$, $\bar{p}_6^{(n)}(0) \equiv \left( p_6^{(n)}(0) / p_1^{(n-5)}(0) - 1 \right) / thv \cdot f$ and $\bar{p}_7^{(n)}(0) \equiv \left( p_7^{(n)}(0) / p_1^{(n-6)}(0) - 1 \right) / thv \cdot f$. In these expressions, $thv \equiv (e^{v} - e^{-v})/(e^{v} + e^{-v})$.

Table-4 Calculated $\bar{p}_i^{(n)}(0)$ of the straight n-string for n=1 to 13 in the external field of Eq.(4)

| n | $\bar{p}_1^{(n)}(0)$ | $\bar{p}_2^{(n)}(0)$ | $\bar{p}_3^{(n)}(0)$ | $\bar{p}_4^{(n)}(0)$ | $\bar{p}_5^{(n)}(0)$ | $\bar{p}_6^{(n)}(0)$ | $\bar{p}_7^{(n)}(0)$ |
|---|---|---|---|---|---|---|---|
| 2 | $1 + \dfrac{e^{-v}}{p_1^{(1)}(0) 2shv}$ | | | | | | |
| 3 | $1 + \dfrac{e^{-v}}{p_1^{(2)}(0) 2shv}$ | 1 | | | | | |
| 4 | $1 + \dfrac{e^{-v}}{p_1^{(3)}(0) 2shv}$ | 1 | | | | | |
| 5 | $1 + \dfrac{e^{-v}}{p_1^{(4)}(0) 2shv}$ | 1 | $\dfrac{2e^{v}}{2chv}$ | | | | |
| 6 | $1 + \dfrac{e^{-v}}{p_1^{(5)}(0) 2shv}$ | 1 | $\dfrac{2e^{v}}{2chv}$ | | | | |
| 7 | $1+\dfrac{e^{-v}}{p_1^{(6)}(0)2shv}$ | 1 | $\dfrac{2e^v}{2chv}$ | $\dfrac{3e^{2v}+e^{-2v}}{(2chv)^2}$ | | | |
| 8 | $1+\dfrac{e^{-v}}{p_1^{(7)}(0)2shv}$ | 1 | $\dfrac{2e^v}{2chv}$ | $\dfrac{3e^{2v}+e^{-2v}}{(2chv)^2}$ | | | |
| 9 | $1+\dfrac{e^{-v}}{p_1^{(8)}(0)2shv}$ | 1 | $\dfrac{2e^v}{2chv}$ | $\dfrac{3e^{2v}+e^{-2v}}{(2chv)^2}$ | $\dfrac{4e^{3v}+4e^{-v}}{(2chv)^3}$ | | |
| 10 | $1+\dfrac{e^{-v}}{p_1^{(9)}(0)2shv}$ | 1 | $\dfrac{2e^v}{2chv}$ | $\dfrac{3e^{2v}+e^{-2v}}{(2chv)^2}$ | $\dfrac{4e^{3v}+4e^{-v}}{(2chv)^3}$ | | |
| 11 | $1+\dfrac{e^{-v}}{p_1^{(10)}(0)2shv}$ | 1 | $\dfrac{2e^v}{2chv}$ | $\dfrac{3e^{2v}+e^{-2v}}{(2chv)^2}$ | $\dfrac{4e^{3v}+4e^{-v}}{(2chv)^3}$ | $\dfrac{5e^{4v}+10+e^{-4v}}{(2chv)^4}$ | |
| 12 | $1+\dfrac{e^{-v}}{p_1^{(11)}(0)2shv}$ | 1 | $\dfrac{2e^v}{2chv}$ | $\dfrac{3e^{2v}+e^{-2v}}{(2chv)^2}$ | $\dfrac{4e^{3v}+4e^{-v}}{(2chv)^3}$ | $\dfrac{5e^{4v}+10+e^{-4v}}{(2chv)^4}$ | |
| 13 | $1+\dfrac{e^{-v}}{p_1^{(12)}(0)2shv}$ | 1 | $\dfrac{2e^v}{2chv}$ | $\dfrac{3e^{2v}+e^{-2v}}{(2chv)^2}$ | $\dfrac{4e^{3v}+4e^{-v}}{(2chv)^3}$ | $\dfrac{5e^{4v}+10+e^{-4v}}{(2chv)^4}$ | $\dfrac{6e^{5v}+20e^v+6e^{-3v}}{(2chv)^5}$ |

The results of table-4 indicate: (1) each column of $\bar{p}_i^{(n)}(0)$ has the same value; (2) from the 2nd column, $p_1^{(n)}(0)=\left[p_1^{(n-1)}(0)e^v+\left(1-p_1^{(n-1)}(0)\right)e^{-v}\right](1+f)/2chv$; (3) for the 3rd to 8th columns, the values of n corresponding to the first terms of the $\bar{p}_{i+1}^{(n)}(0)$ are from top to bottom always odd $(n=2i+1)$, so all $\bar{p}_j^{(n)}(0)$ can be calculated if the general form of $\bar{p}_{i+1}^{(2i+1)}(0)$ is found; (4) the denominator of $\bar{p}_j^{(n)}(0)$ from j=2 to 7 is $(2chv)^{j-1}$, but the numerator is more complex though the following regularities can be found: (i) the coefficients of the exponential are parts of the combinatory number $C_i^l, l=0,\cdots,i$, the multiplying exponential is $e^{(i-1)v}$ in this case and each exponential decreases by a factor $e^{4v}$, e.g., $e^{(i-5)v}$, $e^{(i-9)v}$, $e^{(i-13)v}$; (ii) since each exponential factor of the ordinal $i$ in $(2chv)^i=\sum_{l=0}^{i}C_i^l e^{(i-2l)v}$, decreases by a factor $e^{2v}$, it could be expected that the coefficients of the exponential terms of the numerator of $p_{i+1}^{(2i+1)}(0)$ are $C_i^{2l+1}$, $l=1,\cdots,\text{int}[(i-1)/2]$, where $\text{int}[(i-1)/2]$ is an integer $x$. Then, the relationship $p_{i+1}^{(2i+1)}(0)=p_1^{(i+1)}(0)\left[1+\dfrac{thv}{(2chv)^{i-1}}\sum_{l=0}^{\text{int}[(i-1)/2]}C_i^{2l+1}e^{(i-1-4l)v}f\right]$, $i\geq 1$, can be obtained. In view of this, the recursive equation for $p_i^{(n)}(0)$, n being an arbitrary number, is obtained from table-4 as follows,

$$\begin{cases} p_1^{(1)} = \dfrac{1-f}{2} \\ p_1^{(n)}(0) = \left[p_1^{(n-1)}(0)e^v + \left(1-p_1^{(n-1)}(0)\right)e^{-v}\right]\dfrac{1+f}{2chv},\ n \geq 2 \\ p_{i+1}^{(2i+m)}(0) = p_1^{(i+m)}(0)\left[1+\dfrac{thv}{(2chv)^{i-1}}\sum_{l=0}^{\text{int}[(i-1)/2]}C_i^{2l+1}e^{(i-1-4l)v}f\right],\ i \geq 1,\ m \geq 1 \end{cases} \quad (10)$$

Moreover, the simplified form of Eq.(10) is,

$$\begin{cases} p_1^{(n)}(0) = \dfrac{1}{2}\left[1+\sum_{l=0}^{n-1}(thv)^l f\right],\ n \geq 1 \\ p_{i+1}^{(2i+m)}(0) = \dfrac{1}{2}\left\{1+\left[\sum_{l=0}^{i+m-1}(thv)^l + \dfrac{thv}{(2chv)^{i-1}}\sum_{l=0}^{\text{int}[(i-1)/2]}C_i^{2l+1}e^{(i-1-4l)v}\right]f\right\},\ i \geq 1,\ m \geq 1 \end{cases} \quad (11)$$

Finally,

$$\begin{cases} \delta_1^{(n)}(0) = \dfrac{1}{2}\sum_{l=0}^{n-1}(thv)^l f,\ n \geq 1 \\ \delta_{i+1}^{(2i+m)}(0) = \dfrac{1}{2}\left[\sum_{l=0}^{i+m-1}(thv)^l + \dfrac{thv}{(2chv)^{i-1}}\sum_{l=0}^{\text{int}[(i-1)/2]}C_i^{2l+1}e^{(i-1-4l)v}\right]f,\ i \geq 1,\ m \geq 1 \end{cases} \quad (12)$$

Eq.(12) easily leads to two important conclusions: (1) $\delta_i^{(n)}(0) = f/2$ for $v=0$, the case of an independent molecule Debye relaxation in the double-well $H_0$ of Eq.(1) [9], and (2) $\delta_i^{(n)}(0) = nf/2$ for $v \to \infty$, reflecting the case of the strong correlation limit where the n molecules of an $n$-string behave like a single one. For arbitrary values of $v \equiv V/k_B T$ and $n$, the values calculated for $\delta_i^{(n)}(0)$ are shown in Figs.1 to 3.

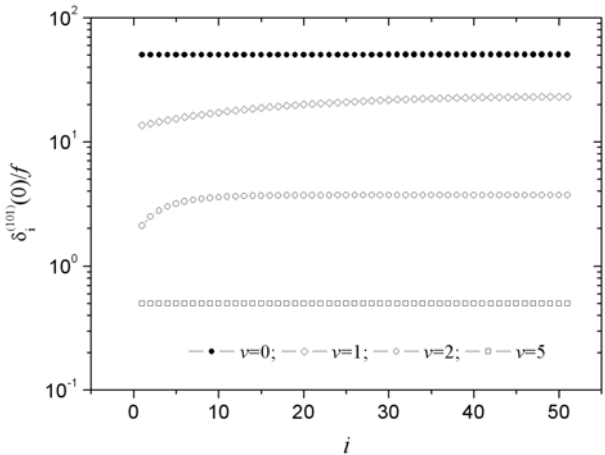

Fig. 1 $\delta_i^{(101)}(0)/f$ as a function of $i$ for a set of positive $v$ values.

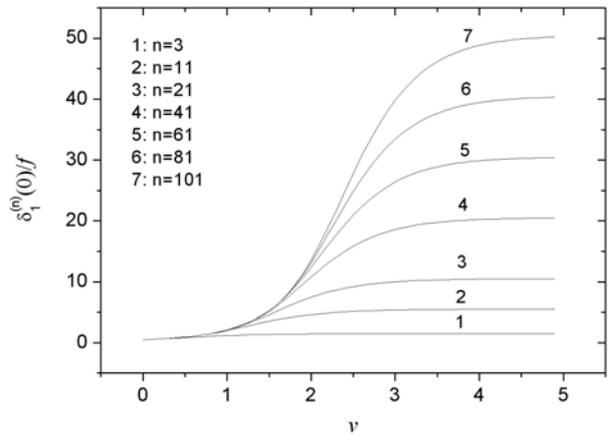

Fig. 2. $\delta_i^{(n)}(0)/f$ as a function of $v$ for a set of $i$ values.

Fig.1 indicates that $\delta_i^{(101)}(0)$ increases with increasing $v$ as a whole. The following points should be highlighted: (1) for small values of $i$ ($i<10$), $\delta_i^{(101)}(0)$ goes up rapidly when $v=1$, but it is almost constant for large values of $i$ ($10<i\leq 51$); (2) $\delta_i^{(101)}(0)$ keeps a slow tendency to increase with increasing $i$ for $v=2$; and (3) the value of $\delta_i^{(101)}(0)$ is nearly independent on $i$ when $v=5$.

To make evident the changes of $\delta_i^{(n)}(0)$ with $v$, values of $\delta_i^{(n)}(0)$ were plotted as a function of $v$ in Fig.2 for n = 3, 11, 21, 41, 61, 81 and 101, respectively. The results indicate that $\delta_i^{(n)}(0)$ is relatively small for $v<1$ but it goes up rapidly as $v$ increases from 2 to 4. However, $\delta_i^{(n)}(0)$ becomes independent of $v$ for $v>5$.

The parameter $\delta_i^{(101)}(0)$ was calculated for negative values of $v$, specifically $v$ = -0.5, -1, -2 and -4, and the pertinent plots $\delta_i^{(101)}(0)$ vs $v$ are shown in Fig.3. The most important difference with the results obtained for $v>0$ is that $\delta_i^{(n)}(0)$ shows a periodic decrease with n, the period being 2 for $v<0$. Specifically, the vibration region increases but the equilibrium value decreases with increasing $|v|$, and the region extends to the whole string though the equilibrium value becomes almost zero.

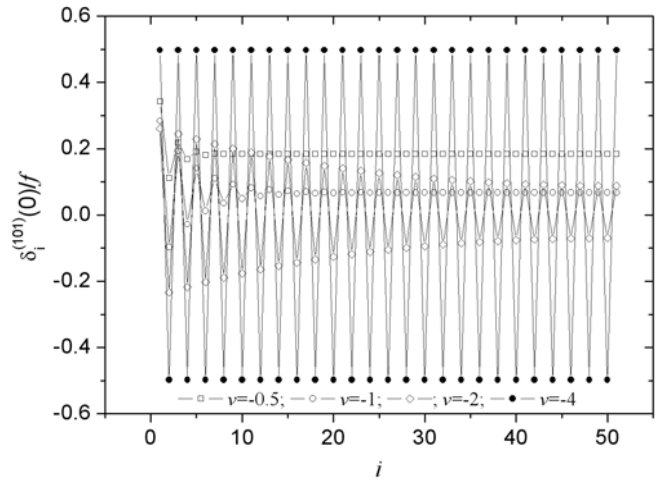

Fig. 3 $\delta_i^{(101)}(0)/f$ as a function of $i$ for a set of negative $v$ values.

Based on the calculated SIC, the general initial condition of the SRE can be calculated. By comparing experiments, we could check whether the model gives a good and unified account of α- and Johari-Goldstein β-relaxation at molecular level. These items will be described in separate publications.

ACKNOWLEDGMENTS

The authors thank Prof. E. Riande for his enlightening discussions. This work was supported by the National Natural Science Foundations of China (Grant No. 10774064 & 10274028), the Key Natural Science Foundation of Xinjiang Educational Department (2008-2010), the Key Natural Science Foundation of Xinjiang Science-Technology Department (2008-2010), and Key Natural Science Foundations of Yili Normal University (2006-2008 & 2007-2009).